\documentclass[12pt,showpacs,preprintnumbers,floatfix]{article}
\usepackage{amsfonts}
\usepackage{amsthm}
\usepackage{amsmath}
\usepackage{amssymb}
\usepackage{graphicx}

\def\be{\begin{equation}}
\def\ee{\end{equation}}
\def\ba{\begin{eqnarray}}
\def\ea{\end{eqnarray}}

\def\bdm{\begin{displaymath}}
\def\edm{\end{displaymath}}
\def\la{~\mbox{\raisebox{-.6ex}{$\stackrel{<}{\sim}$}}~}
\def\ga{~\mbox{\raisebox{-.6ex}{$\stackrel{>}{\sim}$}}~}
\def\bq{\begin{quote}}
\def\eq{\end{quote}}

 at 10truept

\newcommand{\Kbar}{\bar{K}}

\newcommand{\beq}{\begin{equation}}
\newcommand{\eeq}{\end{equation}}
\newcommand{\bea}{\begin{eqnarray}}
\newcommand{\eea}{\end{eqnarray}}
\newcommand{\beqa}{\begin{eqnarray}}
\newcommand{\eeqa}{\end{eqnarray}}

\def\la{~\mbox{\raisebox{-.6ex}{$\stackrel{<}{\sim}$}}~}
\def\ga{~\mbox{\raisebox{-.6ex}{$\stackrel{>}{\sim}$}}~}

\def\ltap{\ \raise.3ex\hbox{$<$\kern-.75em\lower1ex\hbox{$\sim$}}\ }
\def\gtap{\ \raise.3ex\hbox{$>$\kern-.75em\lower1ex\hbox{$\sim$}}\ }
\def\gl{\ \raise.5ex\hbox{$>$}\kern-.8em\lower.5ex\hbox{$<$}\ }
\def\roughly#1{\raise.3ex\hbox{$#1$\kern-.75em\lower1ex\hbox{$\sim$}}}

\begin{document}

\thispagestyle{empty}
\begin{flushright}
November 2012
\end{flushright}
\vspace{1cm}
\begin{center}
{\Large \bf Topological Ghosts: the Teeming of the Shrews}\\

\vspace*{1.25cm} {\large Nemanja Kaloper\footnote{\tt
kaloper@physics.ucdavis.edu} and McCullen Sandora\footnote{\tt
mesandora@ucdavis.edu} }\\
\vspace{.5cm} {\em Department of Physics, University of
California, Davis, CA 95616, USA}\\

\vspace{1.5cm} ABSTRACT
\end{center}
We consider dynamics of spacetime volume-filling form fields with
``wrong sign" kinetic terms, such as in so-called Type-II$^*$ string theories. 
Locally, these form fields are just additive renormalizations of the cosmological constant. They have no fluctuating degrees of freedom.
However, once the fields are coupled to membranes charged under them, their configurations are unstable: by a process analogous to Schwinger pair production the field space-filling flux increases. This reduces the cosmological constant, and preserves the null energy condition, since the processes that can violate it by reducing the form flux are very suppressed. The increase of the form flux implies that as time goes on  the probability for further membrane nucleation {\it increases}, in contrast to the usual case where the field approaches its vacuum value and ceases to induce further transitions. Thus, in such models spaces with tiny positive vacuum energy are ultimately unstable, but the instability may be slow and localized. In a cosmological setting, this instability can enhance black hole rate formation, by locally making the vacuum energy negative at late times, which constrains the scales controlling membrane dynamics, and may even collapse a large region of the visible universe.

\vfill \setcounter{page}{0} \setcounter{footnote}{0}
\newpage

Much of the discourse in inflationary cosmology revolves around the question of how to start inflation. The prevailing idea is that inflation is eternal, without a real beginning, because early on random quantum fluctuations scan through all available phase space, finding the right conditions and depositing the inflaton field at values where its potential rapidly overwhelms its kinetic energy \cite{eternal}. The regions where this doesn't occur either collapse, or produce ``bad" universes, but the total future volume occupied by such regions is overwhelmed by the volume of the universes which did undergo inflation. Classically, the difficulties with ``starting" inflation follow from using General Relativity with matter sources which obey the null energy condition (NEC). These models are all subject to Hawking-Penrose singularity theorems, which dictate the behavior of (semi)classical space-time geometry. If there are sources which might violate NEC, but in ways which do not produce disasters while perhaps helping avoid the singular attractors, one might be able to avoid singularities altogether, as in bouncing cosmologies \cite{prebb,ekpyro,kalol,creminel}, or perhaps even `restart' the universe by facilitating quantum transitions from tame initial states such as an (almost) flat space to a de Sitter space with a large (transient) cosmological constant \cite{markus}. If realized, the latter option could describe the full evolution of the universe completely within the realm of semiclassical theory.

Yet, most sources which violate NEC either can't dominate global evolution \cite{null}, or generically yield to instabilties \cite{ghosts} (that one could attempt to tame by tuning the parameters of the theory \cite{creminel}). On the other hand, there is a candidate for trying to violate NEC without involving any local degrees: a $D$-form field with a {\it negative} kinetic term in $D$-dimensions. Such setups are motivated by allowing for dualities that swap timelike and spacelike directions, as in the so-called Type-II$^*$ theories \cite{hull}, which one can obtain by compactifying timelike directions. It was argued that ghosts may be absent in the full theory. In fact, $D$-forms in $D$-dimensions, by the symmetries of the theory, have {\it no} local degrees of freedom. Their electric and magnetic component coincide, and by local field equations they must be a constant, that measures the space-filling flux \cite{brownteitelboim}. This flux can only change if there are membranes charged under the 4-form U(1) gauge theory, whose nucleation produces bubbles in spacetime, jumping the flux between the interior and exterior of the bubble. When one takes the kinetic term with the normal, `positive' sign, the 4-form flux is locally behaving like an extra positive contribution to the cosmological constant, and the membrane nucleation discharges it, reducing the cosmological constant inside the bubble. This is consistent with no NEC violations. However, with negative kinetic term, the flux is a {\it negative} contribution to the cosmological constant, and so its discharge would {\it increase} the cosmological constant, in effect manifesting NEC violation, but without explicit ghosts. 

The question is, whether such NEC violations really are favored by the dynamics of the theory. In this note we will address this question in a simple theory of a $4$-form coupled to gravity, with a negative kinetic term, and with membranes charged under the $4$-form gauge theory. Using the thin wall limit for the description of the membranes, and imagining that their quantum production goes via the Coleman-De Luccia (CdL) instanton nucleation processes \cite{CdL}, we chart out the possible transition channels between the semiclassical vacua in the theory. We find that starting from a positive cosmological constant, the dominant processes involve nucleation of bubbles which charge up the form flux inside the bubble, analogously to the Schwinger pair production with imaginary charges. So the cosmological constant and the expansion rate of the region inside of the bubble predominantly decrease, instead of increasing, and so on the mean NEC is {\it not} violated. However since the Euclidean action for the membrane nucleation processes is inversely proportional to the flux, the growth of the flux implies that the action becomes smaller and the nucleation rate increases. This is particularly acute during the epoch of eternal inflation; in such a case, the brownian drift would drive the flux towards the regime of Planckian-sized bubbles, which would be totally unsuppressed and would yield a massive breakdown of semiclassical approximation. 
Moreover, after inflation ends, even (an almost) Minkowski space in such theories is generically locally unstable: if the vanishing vacuum energy is realized by a cancellation between a positive bare cosmological constant, and a (large) flux contribution, further nucleation will generically flip the sign of the vacuum energy in the region of space inside the bubble, and with any compressible matter present produce collapse \cite{CdL,abbcol,kallinde,bhs}. 
Nevertheless, while the presence of the $4$-from with negative kinetic term may be detrimental for the theory (it utlimately is), the instabilities may actually be extremely slow, as we shall see from the explicit consideration of the scales controlling the dynamics. Moreover, in a cosmological setting the increase of the flux will stop, because the bubbles that form with a negative vacuum energies inside will collapse and form black holes. So this means there will 
enhanced black hole production at very late times, that would eventually dominate the evolution of the universe. Finally, if they happen to be very large, which may occur if their charge and tension are small, they could even collapse large portions of the visible universe, similarly to 
the examples with local potentials which can be negative, as studied recently in \cite{lindelinder}. 

Let us now turn to the explicit computations illustrating the above. We will work with the action
\ba
S=\frac{M_{Pl}^2}{2}\int_Md^4x\sqrt{|g|}\Bigl(R-2\Lambda\Bigr)+\frac{1}{2}\int_MF\wedge\ast F-M_{Pl}^2\int_{\partial M}d^3\xi\sqrt{|\gamma|}K\nonumber\\+\sigma\int_{\partial M}d^3\xi\sqrt{|\gamma|}-\int_{\partial M}A\wedge\ast F +e\int_{\partial M}A \, , 
\label{action}
\ea
where we include the membrane action with positive tension $\sigma > 0$, to ensure the absence of any explicit local ghosts that could reside on the membrane, and, since the membrane surface is a discontinuity in the gravitational and gauge fields, the gravitational Gibbons-Hawking term and its $4$-form analogue, the Duncan-Jensen \cite{dj} term. These maintain the validity of the standard variational rules on the membrane as a boundary between different spacetime regions.  Here $M$ designates the bulk of the geometry supporting (\ref{action}), while $\partial M$ is the spacetime boundary defined by the membrane wordlvolume(s). Note, that as we define it, the cosmological constant term has dimensions of (mass)$^2$, since we have divided out the Planck scale $M_{Pl}^2$.

We want to find the thin-wall limit of the description of the quantum bubble nucleation, given by the CdL process. This means, that we need to compute the Euclidean action for an instanton configuration, that corresponds to the Euclidean space solution of the theory (\ref{action}) with the boundary conditions on the membrane enforced by the (Euclidean) Israel junction conditions. The Euclidean action is obtained by the substitution $t\rightarrow-it$, which in turn induces the changes in the gauge field sector 
\be A_{0ij} \rightarrow iA_{0ij} \, , ~~~~~~~~~~ F^{\alpha\beta\gamma\delta} \rightarrow 
iF^{\alpha\beta\gamma\delta} \, . 
\label{changes}
\ee
One further needs to bear in mind that under Wick rotation, the charges, as defined by the Gauss law,
\be
e \sim \int_{{\rm timelike~surface~in~}M }{~}^{*}F \, ,
\label{gauss}
\ee
become imaginary as well, $e \rightarrow i e$. This transformation also follows from the fact that the charge is a spatial integral of the timelike component of the current density, and with the standard rules for the gravitational sector. The action (\ref{action}) becomes
$S\rightarrow iS_E$, where
\ba
S_E=\frac{-M_{Pl}^2}{2}\int_Md^4x\sqrt{|g|}\Bigl(R-2\Lambda\Bigr)+\frac{1}{2}\int_MF\wedge\ast F+M_{pl}^2 \int_{\partial M}d^3\xi\sqrt{|\gamma|}K\nonumber\\+\sigma\int_{\partial M}d^3\xi\sqrt{|\gamma|}-\int_{\partial M}A\wedge\ast F +e\int_{\partial M}A\label{S_E} \, .
\label{euclidaction}
\ea

From (\ref{euclidaction}) we can now construct the single instanton solutions, as follows. 
In the bulk, the Euclidean equations of motion are 
\be
G^{\mu\nu} = \Lambda g^{\mu\nu} - \frac{1}{3! M_{Pl}^2} \Bigl(F^{\mu\beta\gamma\delta}F^\nu_{\phantom{1}\beta\gamma\delta} -\frac{1}{8}F^{\alpha\beta\gamma\delta}F_{\alpha\beta\gamma\delta}g^{\mu\nu} \Bigr) \, , ~~~~~~~~~~~~ 
\partial_\mu(\sqrt{g}F^{\mu\nu\sigma\rho})=0 \, .\label{eucleoms}
\ee
The second equation implies $F^{\alpha\beta\gamma\delta}=
\frac{iE}{\sqrt{g}}\epsilon^{\alpha\beta\gamma\delta}$, where $E$ is an integration constant, and the overall imaginary unit is chosen so that its Lorentzian continuation is real. Substituting this solution into the first of Eqs. (\ref{eucleoms}) yields
\be
\Lambda_{eff}=\Lambda - \frac{1}{2 M_{Pl}^2}E^2 \, .
\label{cceff}
\ee
Thus in contrast to the case with positive kinetic term, the flux of a form with a negative kinetic term  reduces the effective cosmological constant. This means, that the bulk solutions are sections of Euclidean spaces with constant curvature, which is the consequence of maximal symmetry, as usual.

When a membrane is nucleated, it yields a jump in the gravitational and gauge fields controlled by the tension and the charge, respectively. The jumps are regulated by the junction conditions on the membrane, which are given by the equations
\be
\Kbar^{ab}- \Kbar \gamma^{ab}=- \frac{1}{2 M_{Pl}^2} \sigma \gamma^{ab}\, , ~~~~~~~~~~~~~
E_{in}=E_{out}\pm e \, ,
\label{junctions}
\ee
The first of Eqs. (\ref{junctions}) is the standard Israel junction condition controlling the jump of the gravitational field across the membrane
with tension $\sigma$. The latin indices take values over the coordinates intrinsic to the membrane, and 
$\gamma_{ab}$ is the intrinsic metric on the membrane. The jump of the extrinsic curvature is given by 
$\Kbar^{ab} = \frac{1}{2} (K^{ab}_{out}-K^{ab}_{in})$, where the labels $(out, ~in)$ refer to the outside and the inside of the spherical membrane and $K^{ab}_{out, in}$ are extrinsic curvatures on the membrane computed relative to the outward-oriented normal. We have used (\ref{euclidaction}) to determine the membrane-localized  stress energy tensor $\bar \tau^{ab} = \frac12 \sigma \gamma^{ab}$.
The second of Eqs. (\ref{junctions}) is the jump of the $4$-form flux if the nucleated membrane carries a single unit of charge $e$. 
Although the sign of the $ 4$-form kinetic term is opposite then its usual, the fact that we are keeping the membrane charge in the action
(\ref{action}) real (by enforcing the kinetic energy sign flip by Eqs. (\ref{changes})), with the conventional definition of the outward normal to the membrane being positive, ensures that the flux changes with a unit charge as in the usual case with positive kinetic terms, as reflected by the signs of $\pm e$ in (\ref{junctions}).  
In fact, requiring that the membrane quantum phase is single-valued in a fixed background flux, one finds that the flux is in fact quantized, in the units of $e$, $E = ne$ (see, {\it eg}. \cite{brownteitelboim,boussopolchinski}). 
Further, when the bulk curvature jumps across the membrane, we need to impose the stress energy conservation $\bar \nabla_\mu \bigl(\Kbar^{ab}- \Kbar \gamma^{ab} + \frac{1}{2 M_{Pl}^2} \sigma \gamma^{ab}\bigr) = 0$, where $\bar \nabla$ is the covariant derivative with respect to the induced metric on the membrane. This equation plays the role of the brane equation of motion. Since 
$\bar \nabla_c \gamma_{ab} = 0$, this reduces to
\be
\bar \nabla_a \Kbar^{ab} = 0 \, .
\label{braneeqs}
\ee

\begin{figure}[thb]
\centering
\includegraphics[height=6cm]{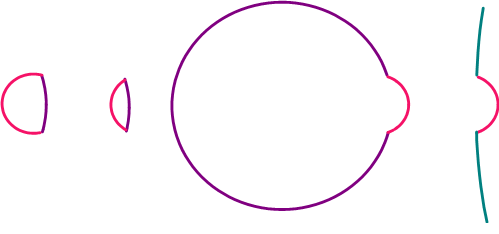}
\caption{low-dimensional depiction of the toenail, hamburger, death star, and volcano instantons, respectively.}\label{lineup}
\end{figure}

So, substituting the formulas for the interior and exterior of the bubble describing maximally symmetric spaces in the first of Eqs. (\ref{junctions}) yields
\be
\pm_{out}\sqrt{1-\frac{\Lambda_{out}\rho^2}{3}}-\pm_{in}\sqrt{1-\frac{\Lambda_{in}\rho^2}{3}}= - \frac{\sigma\rho}{2M_{Pl}^2} \, .
\label{ijc}
\ee
Here $\pm_{in,out}$ correspond to the four different ways we can patch the bulk sections together on 3-spheres of a fixed radius $\rho$, determined by (\ref{ijc}). The conventions follow \cite{brownteitelboim}, where $\pm_{out}=+1$ corresponds to using most of the outer sphere, $\pm_{in}=+1$ corresponds to using the small section of the inner sphere, and so on.  The possible instanton geometries are shown in Fig. \ref{lineup}.
Similarly, the membrane equation of motion (\ref{braneeqs}), after some straightforward algebra gives 
\be
\pm_{out}\sqrt{1-\frac{\Lambda_{out}\rho^2}{3}}+\pm_{in}\sqrt{1-\frac{\Lambda_{in}\rho^2}{3}}=\pm\frac{e}{\sigma}(E_{out} + E_{in}) \, \rho \, .
\label{lorfor}
\ee
This effectively corresponds to the condition that the Lorentz force acting on a higher-codimension object is balanced by the object's tension at the instant of nucleation. The balance condition stems from the fact that the instantons of maximal symmetry are stationary when nucleated in thin wall approximation. Indeed, this precisely follows from the energy conservation between the bulk and the membrane, as discussed in the 
Appendix of \cite{brownteitelboim}. 
The $\pm$ signs on the RHS refer to whether the 4 form field increases or decreases in the interior of the bubble, with the standard convention of the outward normal being positive, as noted above. These fields contribute to the effective cosmological constant negatively, meaning that $\Lambda$ responds conventionally to the choice of sign.  Since the left hand sides of  (\ref{ijc}) and (\ref{lorfor}) only depend on $\Lambda$ and not $E$, the \emph{geometry} of the instantons is unchanged, despite the fact that the field configurations are. 

Let us now explicitly list the candidate solutions of
Eqs. (\ref{ijc}), (\ref{lorfor}). The possible configurations are given in Table (\ref{tableone}):
\begin{table}[hb]
\vskip.4cm
\begin{center}
\begin{tabular}{|c|c|c|c|}
\hline Transition & $(\pm_{out},\pm_{in})$ & Vacuum energies & Nickname \\
\hline
dS$\rightarrow$dS & $(+,+)$ & $\Lambda_{in}<\Lambda_{out}$ &  toenail\\
 & $(-,+)$ & $\Lambda_{in}\gtrless\Lambda_{out}$ &  hamburger\\
 & $(-,-)$ & $\Lambda_{in}>\Lambda_{out}$  & death star\\
\hline
dS$\rightarrow$AdS & $(+,+)$ &  & concave toenail\\
 & $(-,+)$ &  & concave hamburger\\
\hline
AdS$\rightarrow$dS & $(-,+)$ &  &  concave hamburger\\
 & $(-,-)$ &  &  concave toenail\\
\hline
AdS$\rightarrow$AdS & $(+,+)$ & $|\Lambda_{in}|>|\Lambda_{out}|$ & concave volcano\\
 & $(-,+)$ & $|\Lambda_{in}|\gtrless|\Lambda_{out}|$ &  contact lens\\
 & $(-,-)$ & $|\Lambda_{in}|<|\Lambda_{out}|$ & concave volcano\\
\hline
\end{tabular}
\end{center}
\caption{Candidate euclidean solutions: single instanton cases }\label{tableone}
\end{table}

\noindent Note, that for a given set of parameters in the action, specifically $\sigma$ and $e$ characterizing the membrane, the equations (\ref{ijc})-(\ref{lorfor}) determine the radius of the nucleated membrane and the value of the interior flux, given the initial (exterior) solution. This means, that the solutions will not exist for all values of parameters (some will not exist for {\it any} values of the parameters). However, for a fixed $\sigma, e$, some of the above configurations will always exist for a given $\Lambda$, since the background $\Lambda$ scans all the possible values between $\pm ({\rm cutoff})^4$. 

To determine which of these configurations are physically most relevant, i.e. dominant, we need to compute the value of the Euclidean action of an instanton configuration relative to a background (exterior) geometry without an instanton. This difference of the two actions is given by 
\be
B = S_{bulk} + S_{boundary} - S_{bulk}({\rm background})\, .
\label{euclactions}
\ee
The probability to nucleate the membrane is then given, to the leading order, and in the dilute instanton gas approximation, by 
\be
\Gamma\propto e^{-B} \, .
\label{rate}
\ee
This formula is a good approximation when the $O(4)$-invariant solution is the minimum of the Euclidean action, which is can be seen from inspecting (\ref{S_E}), because the only difference relative to the case with standard signs in the action are the $4$-form field strength signs. However, these merely change signs of particular cosmological constant contributions, and the local variations of the matter sector are still positive energy. Thus 
the contributions which violate the $O(4)$ symmetry will be penalized by energy cost, specifically from the contributions involving the membrane tension $\sigma$.

\begin{figure}[t]
\centering
\includegraphics[height=5cm]{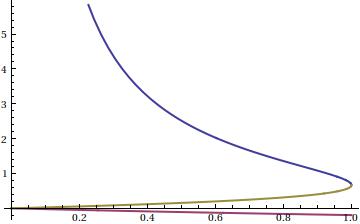}
\caption{Values of the quantity $\Lambda$V($\Lambda$,$\rho$) for various choices of instanton geometry.  The blue curve includes most of the sphere, that is the part including the tropics.  The yellow curve includes just the polar region, and the purple curve is the part of a hyperboloid enclosed within a circle of radius $\rho$.  }
\label{curves}
\end{figure}

In the explicit form, using the bulk equations (\ref{eucleoms}), (\ref{cceff}) and junction conditions
(\ref{ijc}), (\ref{lorfor}), the effective Euclidean action $B$ for any single-instaton configuration of
Table (\ref{tableone}) is
\be
B =-M_{Pl}^2\Lambda_{in}V(\Lambda_{in},\rho)+ M_{Pl}^2 \Lambda_{out}V^c(\Lambda_{out},\rho)+\sigma A(\rho) \, .
\label{exponent}
\ee
Here $V$ is the volume of the interior of the (spherical) membrane outlining the boundary of the instanton configuration, and $V^c$ is the proper volume of the eliminated section of the background which is occupied by the new interior geometry. The individual contributions are evaluated at the extremum value of the radius $\rho$, determined by the junction conditions (\ref{ijc}), (\ref{lorfor}).  The specific formulas for the volumes, for a given $\rho$, depend on the intrinsic bulk geometries. Because the curvatures before and after the transition are different, so are these proper volumes. The factor $A(\rho)$ is the surface area of the membrane, given by $A(\rho)=2\pi^2\rho^3$. Note, that the residual section of the background exterior to the membrane exactly cancels out from $B$. Let us now consider an explicit example.
For instance, taking the transition between two Euclidean de Sitter spaces, and placing the interior with a larger curvature radius inside a smaller `polar cap' portion of the exterior sphere, we find
\be
V(\Lambda,\rho)=\frac{18\pi^2}{\Lambda^2}\bigg(\frac{2}{3}+\frac{1}{3}\big(1-\Lambda\rho^2/3\big)^{3/2}-\sqrt{1-\Lambda\rho^2/3}\bigg) \, .
\label{volumes}
\ee
On the other hand, if a transition existed (as controlled by the junction conditions (\ref{ijc}), (\ref{lorfor})) 
which required that the interior of the membrane had to include the complement of the small polar cap, thus involving the equatorial belt, the expression for the volume (\ref{volumes}) would change to
 $V_{polar+tropics}=24\pi^2/\Lambda^2-V_{polar}$. Similar expressions cover other possible cases of Table (\ref{tableone}). The values of 
 $\Lambda V$ for various configurations are depicted in Fig. (\ref{curves}).

We can obtain further insight into the dynamics of these processes by determining the radius of the bubbles at the instant of nucleation. 
Solving (\ref{ijc}) for the radius, we find
\be
\rho=\frac{\sigma/M_{Pl}^2}{\sqrt{-\frac{4}{9}\Lambda_{in}\Lambda_{out}+(\frac{\Lambda_{in}+\Lambda_{out}}{3}+\frac{\sigma^2}{4M_{Pl}^4})^2}} \, .
\label{radiusinst}
\ee
There are two main dynamical regimes, depending on whether the scale of the tension of the bubble dominates over the bulk vacuum energy scale, or not. The first regime occurs when the tension scale dominates, $\sigma > M_{Pl}^2 \sqrt{\Lambda_{in}+\Lambda_{out}}$. In this case the radius of the instanton is completely controlled by the tension, 
\be
\rho = 2\frac{M_{Pl}^2}{\sigma} \,.
\label{tenrad}
\ee
This limit is relevant for understanding bubble production in a very low curvature environments, with a tiny membrane charge. It follows from the Israel junction condition (\ref{ijc}) when charges are subdominant to the tension. 
Direct inspection shows that this configuration is dominant in the case of $(-,+)$ transitions.  
This explains $\rho \propto \frac{1}{\sigma}$, since in this case a
smaller bubble radius actually corresponds to a larger change in geometry. It follows from ``energy balance" equation, which 
requires $\rho$ to be smaller for larger $\sigma$ in the thin wall limit, and vice-versa.
However, the action for such instantons is $B\sim M_{Pl}^6/\sigma^2$, and when we limit the tension to not ever exceed the Planck scale, a reasonable assumption given that the tension of a thin membrane in reality represents the limit of the local interaction scale of the objects condensing to form the membrane, the transitions are 
heavily suppressed\footnote{A (very remote) hance that there are membranes with superplanckian tension, which would yield rapid channels for transitions, is interesting but requires an explicit UV completion to analyze in more detail. That is far beyond the scope of the present work.}.

The other case is the decoupling limit of gravity, where the bubble nucleation is controlled by the locally Lorentzian dynamics. Taking $M_{Pl}\rightarrow\infty$, the charge term dominates and the radius is given by
\be
\rho = \frac{3\sigma}{ne^2} \, ,
\label{radcharge}
\ee
which agrees with the (almost) flat space analysis of transitions in \cite{boussopolchinski,cascade}, as expected for very small bubbles. The inclusion of 
a finite $M_{Pl}$ implies that this expression remains valid when the tension is small, 
$\frac{\sigma^2}{M_{Pl}^4} < \frac{(\Delta \Lambda)^2}{\Lambda_{in} + {\Lambda}_{out}} \simeq \frac{\langle E \rangle^2 e^2}{M_{Pl}^4(\Lambda_{in} + \Lambda_{out})} $, where $\langle E \rangle$ is the mean $4$-form flux on the membrane. This condition is generically satisfied when the environmental cosmological constant is large, since then the change $\Delta \Lambda$ is smaller than $\Lambda$. When the environmental cosmological constant outside of the bubble is small, $\Delta \Lambda \la (\Lambda_{in} + \Lambda_{out})$, and as long as 
$\frac{\sigma^2}{M_{Pl}^4} < \Delta \Lambda \sim \frac{\langle E \rangle e}{M_{Pl}^2} $ or therefore $\sigma < \sqrt{\langle E \rangle e} M_{Pl} \simeq \sqrt{n} eM_{Pl}$ this approximation remains valid. 
Further, these bubbles are always strictly smaller than the horizon size;  when the environmental cosmological constant is large, $H^2 \rho^2 \simeq \frac{\sigma^2 \Lambda_{out}}{n^2 e^4} \la \frac{\Lambda_{out}}{\Lambda_{in} + \Lambda_{out}}<1$, and when it is small,
$\rho < \frac{M_{Pl}}{e}$, which is easily much shorter\footnote{Violating this in our universe would require membranes with 
the scale of charge smaller than $10^{-3}$ eV, a highly unrealistic situation.} than $H^{-1}$. While it may be arranged to have very large bubbles by fine tuning the parameters of the theory, and picking a special background, we will mainly ignore these nonlinear regimes are some since the salient features of the dynamics can be seen in the small bubble limit. In all these cases the dominant transition is mediated by the ``toenail" instanton, as in the case of the usual kinetic term sign - it has the smallest positive action. This means, that the vacuum energy continues to decrease by membrane nucleation. For small bubbles, the volume terms reduce to $V(\rho)=\pi^2\rho^4/2$, and so the Euclidean action (\ref{exponent}) simplifies to 
\be
B=- \frac{\pi}{2} \Delta E^2 \rho^4+2\pi\sigma\rho^3=\frac{27\pi^2}{2}\frac{\sigma^4}{n^3e^6} \, .
\label{smallB}
\ee
This remains positive because the area contribution dominates the Euclidean action (\ref{exponent}), as in the standard case. This can be understood by an analogy with the problem of determining circular orbits in a central potential. As in that case, the `angular momentum conservation' (our Eq. (\ref{lorfor}), which fixes the numerical value of $\Lambda_{in}$ once $\Lambda_{out}$ and $\rho$ are given), reduces the problem to a one-dimensional one, controlled by the `potential' (\ref{exponent}), whose extremum is given by Eq. (\ref{ijc}). The instanton should be a {\it maximum} of the Euclidean action, i.e. our `potential' $B$ (\ref{exponent}), and since the area term in (\ref{exponent}) is strictly positive, as we demand $\sigma$ to be nonnegative to guarantee the absence of any local ghost modes, the bulk term must be negative for a maximum to exist. 
Thus the `potential'  (\ref{exponent}) - in this limit (\ref{smallB}) - will be positive. Because the area term grows more slowly with $\rho$, it must dominate the `potential' $B$ at the maximum. 

Similar reasoning can be used to understand other transitions. A summary of solutions is given in the Table (\ref{tabletwo}).
\begin{table}[hb]
\begin{center}
\begin{tabular}{|c|c|c|}
\hline Transition & Nickname & Bounce action Ê\\
\hline
dS$\rightarrow$dS & toenail & $B>0$ (minimal action, dominant process) \\
& hamburger & $B>0$ \\
& death star & $B>0 $ \\
\hline
dS$\rightarrow$AdS & concave toenail & $B>0$ (minimal action, dominant process) \\
& concave hamburger & $B>0$ \\
\hline
AdS$\rightarrow$dS & concave hamburger & $B=+\infty$ \\
&concave toenail & $B=+\infty$ \\
\hline
AdS$\rightarrow$AdS &concave volcano & $B>0$ \\
& contact lens & $B=+\infty$ \\
& concave volcano Ê& $B=+\infty$ \\
\hline
\end{tabular}
\end{center}
\caption{Signs of the instanton actions}\label{tabletwo}
\end{table}
Since the dynamics obeys the same extremality conditions as in the theory with standard sign kinetic terms, the instanton actions are always positive.
Transitions other than $(+,+)$ have an action dominated by the creation or destruction of a large portion of Euclidean (Anti) de Sitter space, with
$B\propto M_{Pl}^2/\Lambda$, and so these actions are huge. In fact they are infinite for the cases with $AdS$ topology change, and always positive. Further, away from the small bubble limit the area term still dominates, by the arguments involving the bounce action being a maximum of the Euclidean action, which implies that the volume contribution is still of the order estimated above.  However, configurations that have bubbles close to the radius of curvature of the universe are somewhat pathological, as they would require a precise tuning of parameters for the denominator of (\ref{radiusinst}) to be small. We will ignore them since they are only possible as transients.

The main difference between the theory with the negative $4$-form kinetic terms and the standard lore is that here the $4$-form fluxes predominatly charge up during transitions. This is a behavior that indicates an instability, but still the transition is described by a positive Euclidean action, which implies that the instability is slow. The reason behind this can be understood as follows. While the action of the flux sector is by itself negative, favoring the flux increase, rather than discharge, to facilitate such transitions one must make membranes to carry away the charge difference. This follows from unbroken gauge invariance. But since the membranes have positive tension, their creation costs a large positive contribution to the action, which in fact dominates at the extremum point, by symmetry. Thus the total Euclidean action remains positive. So, in principle, the instability may end up being very slow. The $4$-form with a negative kinetic term in a sense behaves like a ghost, by reducing vacuum energy towards negative values, which at least in principle are not bounded from below.

In the course of evolution, as the flux increases, the transition rates go up. Indeed, if we consider the most interesting limit $\sigma < \sqrt{n} e M_{Pl}$, we see that in the course of evolution $n$ is driven to larger and larger values, while transitions remain suppressed as dictated by, eg. Eq. (\ref{smallB}). However, the larger $n$ gets, the smaller $B$, and so the less suppressed transitions will be. Simultaneously, since the radius depends inversely on the amount of flux, with each successive nucleation the bubbles will be progressively smaller,  eventually reaching the Planck length. This in fact defines the true UV cutoff of the theory describing the bubble nucleations, and the effective description of models with a negative $4$-form kinetic terms. When the bubble radius is Planckian, the corresponding flux is
\be
n_*=\frac{3M_{Pl}\sigma}{e^2} \, ,
\label{uvflux}
\ee
and so the action is 
\be
B_*=\frac{\pi^2}{2}\frac{\sigma}{M_{Pl}^3} \, .
\label{uvaction}
\ee
This is small for sub-Planckian tensions (actually, it's already tiny for a tension scale smaller than about a tenth of the Planck scale) and the subsequent transitions are completely unsupressed. Once the dynamics reaches this stage -- obviously, we are taking the effective environmental cosmologival constant to still be positive and significant - implying $\Lambda_{bare} \gg \sigma/M_{Pl}$  -- the background becomes completely unstable, `evaporating' away by a swarm of bubbles that are rapidly, and haphazardly, reducing the local values of the cosmological constant (although it may still remain positive in the interior, depending on the ratio of $\Lambda_{out}$ and $e^2/M^2_{Pl}$). What this actually means is that when the evolution reaches this point the (semi)classical description of spacetime breaks down because the quantum instability driven by the membrane nucleation becomes uncontrollable, even if the background curvature is very low.

This stage may take a very long time to reach. If the initial flux of the $4$-form with a negative kinetic term is close to zero (in which case $n \sim 1$ in the Eq. (\ref{smallB})), it may take a very long time before the flux charges up to the point where the quantum effects totally destabilize the background. 
If one attempts to build a phenomenological description of a universe based on a theory with such modes, in principle there will be other channels for lowering the vacuum energy. However, in the framework of eternal inflation, a presence of a $4$-form with a negative kinetic term will definitely be exploited. So as long as the initial cosmological constant, while sub-Planckian, is larger than $\sigma/M_{Pl}$, the stage with the critical value of the flux $n_*$ will definitely be reached, and such regions will in fact statistically proliferate. Inside these regions, self-reproduction processes will run rampant, since the action $B_*$ will be tiny. Hence, a $4$-form with a negative kinetic term will drive eternal inflation into a UV catastrophe.

This may be avoided in the regions of space where other vacuum energy relaxation mechanisms kick in sooner, reducing the environmental cosmological constant below $\sigma/M_{Pl}$. Subsequent bubbles of our $4$-form would produce regions of the Universe with a large negative cosmological constant in their interior. They would rapidly collapse \cite{CdL,abbcol,kallinde,bhs}, within a time $\tau \sim \rho$ after bubble nucleation \cite{abbcol}. So these processes would produce lots of `primordial'-like black holes in a {\it very late} universe. The size of these black holes would be given by their mass, $R_{BH} \sim M_{BH}/M_{Pl}^2$. If they were really small, they'd evaporate quickly. If they aren't so small, they would stick around for a rather long time. Either way, they would inject extra energy into the universe. To see how much, we can first estimate the mass by the surface energy of the bubble at the time of nucleation (the (negative) energy density inside being compensated by the (negative) curvature of the bubble), 
\be
M_{BH} \sim A_{bubble} \sigma  \simeq \frac{\sigma^3}{n^2 e^4}  \, .
\label{bhmass}
\ee
These black holes generically are small, their Schwarzschild radius being 
$R_{BH} \sim M_{BH}/M_{Pl}^2 \simeq \frac{\sigma^3}{n^2 e^4 M_{Pl}^2}$, and so it is maximized for $n \sim 1$. Since in this regime 
$\sigma \sqrt{n} eM_{Pl}$, it follows that $R_{BH} < \frac{M_{Pl}}{\sqrt{n}e}$, so that it would take absolutely tiny charges to get these black holes to be large. For example, in the present universe, since $H_0 R_{BH} < \frac{H_0 M_{Pl}}{\sqrt{n} e} < \frac{H_0 M_{Pl}}{e}$ it would take a charge scale
smaller than $10^{-3}$ eV to ever even have a chance of making a large enough bubble to make such a large black hole (as we already noted in footnote 2). 

To estimate their abundance, we need the bubble nucleation rate, which we can estimate by the standard expression \cite{jaumeg}
\be
\frac{\Gamma}{V} \sim \frac{e^{\zeta'_R(-2)}}{4} ~ \bigg(\frac{\sigma}{M_{Pl}H^2}\bigg)^2H^4  ~ \exp({-\frac{27\pi^2}{2} \frac{\sigma^4}{n^3e^6}}) \, ,
\label{ratetotal}
\ee
when the nucleation is relatively slow. The Riemann function bit,  $\propto \zeta'_R$, is irrelevant. For a fixed charge and tension, it's obvious that the probability rate for nucleating a bubble increases as $H$ decreases. So the largest
production rate for individual bubbles will occur in the present Universe, where the number of bubbles nucleated inside the Hubble volume and over a Hubble time (now) is
\be
{\cal N} \sim \frac{\Gamma}{H_0^4 V} \sim \bigg(\frac{\sigma}{M_{Pl}H_0^2}\bigg)^2 ~ \exp({-\frac{27\pi^2}{2} \frac{\sigma^4}{n^3e^6}}) \, ,
\label{problate}
\ee
Their total energy density inside a Hubble volume is $\rho_{DM} \simeq {\cal N} M_{BH} H_0^3$. This can't exceed the present energy density of the Universe, of the order of $M_{Pl}^2 H_0^2$ .  So, substituting in the formulas (\ref{bhmass}) and (\ref{problate}) we must demand 
\be
\Bigl(\frac{\sigma}{M_{Pl} H_0^2} \Bigr)^2 \, \exp(- \frac{27\pi^2}{2}  \frac{\sigma^4}{n^3 e^6}) 
\frac{\sigma^3}{n^2 e^4} ~ H_0 =  \exp(- \frac{27\pi^2}{2}  \frac{\sigma^4}{n^3 e^6}) ~
\frac{\sigma^5}{n^2 e^4 H_0^3 M_{Pl}^2} \la M_{Pl}^2 \, ,
\label{ineq}
\ee
to satisfy the observational bounds, along with the inequalities $\sigma < \sqrt{n} e M_{Pl}$ and $e > H_0 M_{Pl}$.
This inequality provides a rough bound on the membrane tension in the units of its charge. The bound still depends on the number of flux units $n$ in the late universe. We can estimate the flux as follows.  Again taking the universe to emerge from eternal inflation, which in some region ended before 
driving our $4$-form's flux to the `Planckian' value $n_*$, we can estimate the most likely value of $n$ by its random diffusion from the origin, which 
gives $n_{typical} \sim \sqrt{n_*} \simeq \sqrt{\frac{M_{Pl} \sigma}{e^2}}$. Substituting this in the inequality above yields
\be
 \exp(- \frac{27\pi^2}{2}  \frac{\sigma^{5/2}}{e^3 M_{Pl}^{3/2}}) ~ \frac{\sigma^4}{e^2 M_{Pl}^2} \la H_0^3 M_{Pl}^3 \, .
\label{ineq2}
\ee
The expression on the right hand side is the critical density of the universe to the power 3/2. So, in the units of Planck mass, it is $10^{-120}$. If we express the left hand side in terms of the quantitity $x = \frac{\sigma^{5/2}}{e^3 M_{Pl}^{3/2}}$, we find
$e^{-133 x} ~ x^{8/5} ~ (\frac{e}{M_{Pl}^2})^{14/5} \simeq 10^{-58 x} ~ x^{8/5} ~ (\frac{e}{M_{Pl}^2})^{14/5} \la 10^{-120}$. To satisfy the inequality, we must either {\it i)} take $e/M_{Pl}^2 \ll 1$; {\it ii)} take $x \ll 1$; or {\it iii)} take $ x \ga {\cal O}({\rm few})$. The first case corresponds to the 
$4$-form decoupling limit, where the membranes aren't emitted any more and the flux is approximated by a constant; that's trivial and we shall ignore this limit. In the second case, we would lose any suppression from the instanton action, and it would imply that the semiclassical description of  
membrane nucleation breaks down, indicating a massive quantum instability of the theory. Hence we will ignore this option too. This leaves us with requiring $x \ga {\cal O}({\rm few})$, which means that the black hole abundance is suppressed by the exponential. This implies that 
$\sigma^5 \ga {\cal O}(10) \, e^6 M_{Pl}^3$. The inequality $\sigma < \sqrt{n} eM_{Pl}$ translates to $e^2 M_{Pl}^5 \ga \sigma^3$ after substituting in $n \sim n_{typical}$. These two inequalities first imply $e \la M_{Pl}^2$, which in fact is expected in general, and then `sandwich' in the values of the tension below the curve $e^{2/3}$ and $e^{6/5}$, when we express all the dimensional quantities in the units of the Planck mass. Clearly, since $e <1$ in these units, solutions do exist. Thus, it at least seems possible that such black holes may in fact form in the late universe. Since they will be proliferated at later times even more efficiently, they, or the products of their decay, would eventually dominate over the primordially created matter.

Finally, we also note the extremely curious, and rather ominous possibility of creating very large black holes, when 
$\sqrt{e} \simeq H_0 M_{Pl} \sim 10^{-3}$ eV. While this is certainly very fine-tuned, it's not a priori excluded. In this case, our two inequalities place a bound on the tension, $10 ~ {\rm TeV} \ga \sigma^{1/3} \ga {\rm keV}$. On the other hand, if a bubble with such a charge were nucleated, it would take the time comparable to the age of the Universe to form a black hole, implying that for a long time its wall would be visible to observers in the late Universe. So the tension must also be bounded by the absence of domain wall distortions in the CMB, which is currently estimated to be at 
$\sigma^{1/3} \la $ 100 keV \cite{zeldovich}; so really the tension is bounded by $100 ~ {\rm keV} \ga \sigma^{1/3} \ga {\rm keV}$.
If it saturates the upper limit\footnote{For our assumed value of the $4$-form flux,$n \sim n_{typical} \gg 1$ this would make the action (\ref{smallB}) large, and so the nucleation rate (\ref{ratetotal}) very small. However, a change of $n$ by only a few orders of magnitude can compensate this, and make the nucleation rate significant.}
the tension could at least in principle leave small imprint in the CMB, close to the detectable limits, inducing directionality in the sky. In this way, one might end up with a mechanism for generating the so-called `axis of evil' \cite{magoo}, which has been widely discussed recently. The name would be quite appropriate in this case, since detection of such an effect might be a grave portent of an impending cataclysm, the local collapse of the 
region of the Universe into a black hole, with consequences similar to those studied for collapse driven by a potential of a scalar field turning negative \cite{lindelinder}. 

To summarize, we have found that a theory with a $4$-form with a negative kinetic term has an instability, which involves nucleation of bubbles which charge up the form flux inside the bubble. This preserves NEC, since the cosmological constant inside the bubble decreases, increase of the flux implies that the nucleation rate increases. Even almost flat spaces in such theories are locally unstable. Further, during eternal inflation the growth of flux would yield a regime of Planckian-sized bubbles, with small Euclidean action, which means that the theory would run away from the semiclassical regime.
However, in the regions where eternal inflation might end before this stage is reached, the instabilities may actually be extremely slow, and so the theory may pass observational bounds. In a late universe, any such bubbles that form would have negative vacuum energies inside and would collapse into black holes. This yields an enhanced black hole production at very late times, that would eventually dominate the evolution of the universe. A particularly ominous possibility is that these black holes happen to be very large, in which case they could collapse large portions of the visible universe.

\vskip.5cm

{\bf \noindent Acknowledgements}
 
\smallskip

We thank Renata Kallosh, Matt Kleban, Lorenzo Sorbo and especially Andrei Linde for useful discussions. This work is
supported in part by the DOE Grant DE-FG03-91ER40674.

\end{document}